\documentclass[11pt,a4paper,useAMS,usenatbib]{emulateapj}
\usepackage{epsfig}
\usepackage{amsmath}
\usepackage{natbib}

\begin{document}

\title{\textit{Chandra} and \textit{ROSAT} observations of Abell 194: detection of an X-ray cavity and mapping the dynamics of the cluster}

\author{\'Akos Bogd\'an\altaffilmark{1}, Ralph P. Kraft\altaffilmark{1}, William R. Forman\altaffilmark{1}, Christine Jones\altaffilmark{1}, Scott W. Randall\altaffilmark{1}, Ming Sun\altaffilmark{2}, Christopher P. O'Dea\altaffilmark{3}, Eugene Churazov\altaffilmark{4}, and Stefi A. Baum\altaffilmark{3}}
\affil{\altaffilmark{1}Smithsonian Astrophysical Observatory, 60 Garden Street, Cambridge, MA 02138, USA}
\affil{\altaffilmark{2}Department of Astronomy, University of Virginia, P.O. Box 400325, Charlottesville, VA 22901}
\affil{\altaffilmark{3}Rochester Institute of Technology, 84 Lomb Memorial Drive, Rochester, NY 14623, USA}
\affil{\altaffilmark{4}Max-Planck-Institut f\"ur Astrophysik, Karl-Schwarzschild-str. 1, 85741 Garching bei M\"unchen, Germany}
\email{E-mail: abogdan@head.cfa.harvard.edu}

\shorttitle{\textit{CHANDRA} AND \textit{ROSAT} OBSERVATIONS OF ABELL 194}
\shortauthors{BOGD\'AN ET AL.}

\begin{abstract}
Based on \textit{Chandra} and \textit{ROSAT} observations, we investigated the nearby poor cluster Abell 194, which hosts two  luminous radio galaxies, NGC547 (3C 40B) and NGC541 (3C 40A). We demonstrated the presence of a large X-ray cavity ($r\sim34$ kpc) formed by the giant southern radio lobe arising from 3C 40B in NGC547.  The estimated age of the cavity is  $t=7.9\times10^7$ years and the total work of the AGN is $3.3\times 10^{59} \ \rm{erg}$, hence the cavity power is $P_{cav}=1.3 \times 10^{44} \ \rm{erg \ s^{-1}}$. Furthermore, in the \textit{Chandra} images of NGC545 and NGC541 we detected sharp surface brightness edges, identified as merger cold fronts, and extended tails. Using the  pressure ratios between inside and outside the cold fronts we estimated that the velocities of NGC545 and NGC541 correspond to Mach-numbers of $M=1.0^{+0.3}_{-0.5}$ and $M=0.9^{+0.2}_{-0.5}$, respectively. The low  radial velocities of these galaxies relative to the mean radial velocity of Abell 194 imply that their motion is oriented approximately in the plane of the sky. Based on these and earlier observations, we concluded that NGC545 and NGC541 are falling through the cluster, whose center is NGC547, suggesting that Abell 194 is undergoing a significant cluster merger event. Additionally, we detected 20 bright  X-ray sources around NGC547 and NGC541, a surprisingly large number, since the predicted number of resolved LMXBs and CXB sources is $2.2$ and $4.1$, respectively. To explain the nature of additional sources, different possibilities were considered, none of which are satisfactory. We also studied the origin of  X-ray emission in Minkowski's Object, and concluded that it is most likely dominated by the population of HMXBs rather than by hot diffuse ISM. Moreover, in view of the galaxy dynamics in Abell 194, we explored the possibility that the starburst in Minkowski's Object was triggered by its past interaction with NGC541, and concluded that it may be a viable path.
\end{abstract}


\section{Introduction}
The essential element of the   cooling flow problem is that in the centers of galaxy clusters the intracluster medium (ICM) should cool on a timescale much shorter than the Hubble time \citep{fabian94}. However, observational evidence from \textit{Chandra} and \textit{XMM-Newton} revealed that only relatively small amounts of gas cools to low temperatures \citep{david01,peterson03,kaastra04}. The most commonly accepted resolution of this problem is the (re)heating of the ICM by the central active galactic nucleus (AGN) \citep[e.g.][]{churazov01,peterson06}. In galaxy clusters the most direct signature of the interaction of AGN outbursts and the ICM is the detection of X-ray cavities \citep[e.g.][]{boehringer93,forman05}. Outbursts from powerful radio sources can significantly increase the entropy of the ICM thereby inflating the gas distribution and reducing the gas density. Hence at the position of radio lobes a decrement in the X-ray surface brightness becomes observable.

\begin{figure*}[t]
\begin{center}
    \leavevmode\epsfxsize=15.8cm\epsfbox{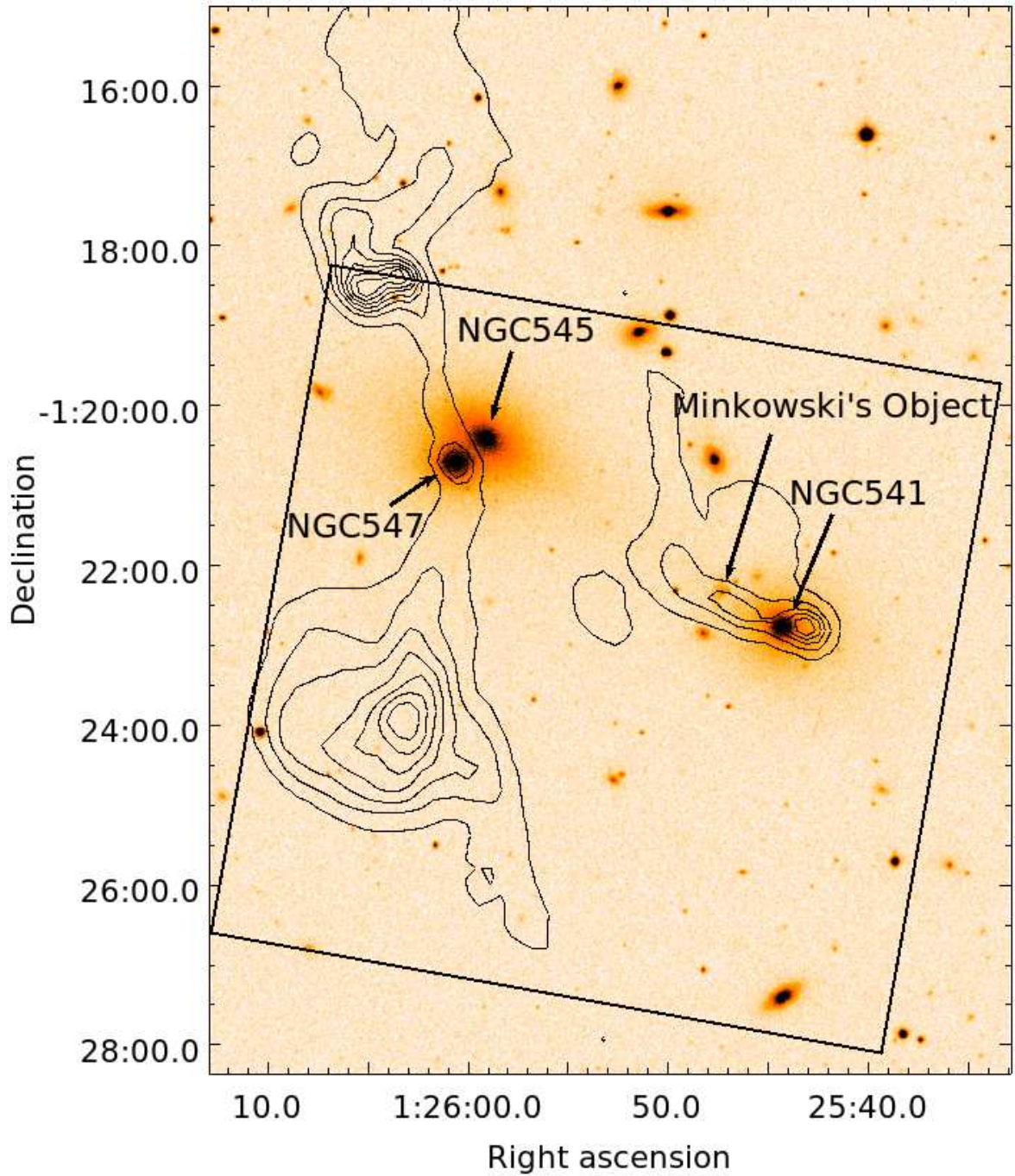}
    \caption{DSS $R$-band image of the central $216 \times 288$ kpc region of Abell 194. Overploted are the intensity levels of the $1.4$ GHz VLA image. The rectangular region shows the field-of-view of the \textit{Chandra} ACIS-S3 detector. The bright radio source, 3C 40B, is centered on NGC547, whereas 3C 40A is associated with NGC541. Minkowski's object, which is believed to be a region of jet induced star formation, is visible approximately $1\arcmin$ to the northeast of NGC541.}
\label{fig:dss_image}
\end{center}
\end{figure*}

\begin{figure*}
  \begin{center}
    \leavevmode
       \epsfxsize=8.5cm\epsfbox{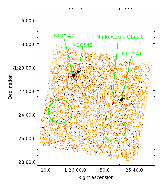}
\hspace{0.75cm} 
      \epsfxsize=8.5cm\epsfbox{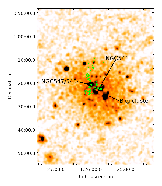}
      \caption{\textit{Left:} $0.98\arcsec$ Gaussian smoothed \textit{Chandra} ($0.3-2$ keV) ACIS-S3 image of the central $216 \times 288$ kpc region of Abell 194. The overplotted two circular regions with $21.6$ kpc ($1\arcmin$) radius were used to study the surface brightness at the position of the cavity (solid circle) and in a neighboring region (dashed circle).  \textit{Right:} $45\arcsec$ Gaussian smoothed \textit{ROSAT} ($0.5-1.5$ keV) PSPC image of a $1.08 \times 1.44$ Mpc region around Abell 194. Overlaid are the intensity levels of the $1.4$ GHz VLA image. Besides the bright cluster galaxies of Abell 194 outstanding is the background cluster at $z\approx0.15$ \citep{mahdavi05}.}
     \label{fig:abell194}
  \end{center}
\end{figure*}

Abell 194, shown in Fig. \ref{fig:dss_image}, is a poor cluster at a redshift of $z=0.018$, and hosts two luminous radio sources, 3C 40B (wide angle tail radio galaxy  -- WAT), associated with NGC547, and 3C 40A (narrow angle tail radio galaxy  -- NAT), associated with NGC541 \citep{odea85}. The jet emanating from the latter is believed to be responsible for triggering star-formation in Minkowski's Object \citep{breugel85,brodie85}. The extended H-alpha emission from 3C 40B \citep[e.g.][]{baum88} is associated with a dust disk and diffuse UV emission \citep{allen02}, the latter likely due to star formation. The nucleus is  not detected at 11.8 microns \citep{wolk10} and the optical emission line spectrum is low excitation \citep{buttiglione09,buttiglione10}, which indicate that the AGN accretion disk is faint -- consistent with a low accretion rate. 

Abell 194 is one of the most striking ``linear'' clusters of galaxies, as its galaxy distribution  and X-ray emission are both linearly elongated along the northeast-southwest direction \citep{chapman88,nikogossyan99,jones99}. The optical bridge between NGC545/NGC547 and NGC541 \citep{croft06} indicates past and/or recent interactions, thereby suggesting that Abell 194 is not a relaxed cluster. As a consequence of the linear distribution of Abell 194, it has not been clear whether any of the massive radio galaxies are in the center of the cluster, and whether  Abell 194 is undergoing a major merger event.  Recently, \citet{sakelliou08} investigated the cluster relying on \textit{XMM-Newton} and radio observations. The \textit{XMM-Newton} data did not reveal significant stripping of the intersellar medium of the massive galaxies, and the bending of the radio jets suggested that both NGC547 and NGC541 are moving with subsonic velocities on the order of few hundred  $ \rm{km \ s^{-1}}$. Thus, \citet{sakelliou08} concluded that the  central region of Abell 194 is relatively quiescent and is not suffering a major  merger event. 

Hereby we present the results of two \textit{Chandra} pointed observations supplemented with an archival \textit{ROSAT} pointing and  complemented with $1.4$ GHz Very Large Array (VLA) data to study the connection between the radio jet originating from 3C 40B and the X-ray surface brightness of the ICM. Furthermore, we aim to study the surface brightness distribution of the hot interstellar medium (ISM) of the massive early-type galaxies in Abell 194 relying on the superb angular resolution of \textit{Chandra}, thereby unveiling the  dynamics of the cluster. Additionally, we also investigate the origin of the population of bright resolved X-ray sources around the massive early-type galaxies located in Abell 194. Throughout the paper we assume $H_0=71 \ \rm{km \ s^{-1} \ Mpc^{-1}}$, $ \Omega_M=0.3$, and $\Omega_{\Lambda}=0.7$ ($1\arcsec=0.36$ kpc at $z=0.018$). The Galactic column density towards Abell 194 is $N_H=3.78 \times 10^{20} \ \rm{cm^{-2}}$ \citep{dickey90}. All errors in the paper represent $1\sigma$ uncertainties, unless otherwise specified.

The paper is structured as follows: in Sect. 2 we introduce the analyzed data and describe its reduction. In Sect. 3 we demonstrate the existence of a cavity, and compute the cavity power. The cold fronts in NGC545 and NGC541 and the dynamics of Abell 194 are studied in Sect. 4. The population of bright resolved sources around NGC541, NGC545, and NGC547 is investigated in Sect. 5. The X-ray emission from Minkowski's Object is studied in Sect. 6. Finally, we conclude in Sect. 7.

\section{Data reduction}
\subsection{Chandra}
\label{sec:chandra}
The central part of Abell 194 was observed by two \textit{Chandra} observations on 2007 September 3 for 9.2 ks (Obs ID: 9583) and on 2007 September 7 for 65.7 ks (Obs ID: 7823) using the ACIS-S detector  (Fig. \ref{fig:abell194} left panel). The data were processed with CIAO 4.3 and CALDB 4.4.2. 

The main steps in the  data analyis agree with those outlined in \citet{bogdan08}. After filtering flare contaminated time intervals, the total remaining exposure time was 72.3 ks. The data were combined by projecting the shorter observation into the coordinate system of Obs ID 7823. Point sources were detected with the CIAO wavdetect tool in three separate energy bands: in the soft (0.5-2 keV), full (0.5-8 keV), and hard (2-8 keV) bands. The scales on which we searched for sources were the  $\sqrt2$-series from 1.0 to 8.0, all other parameters of the wavdetect tool were left at default values. The resulting source lists were combined, which was used to study the population of bright X-ray sources and to mask out the point sources for further study of the diffuse emission. 

To correct for vignetting and to estimate source detection sensitivities, exposure maps were produced using a power law model with slope of  $\Gamma=1.56$, typical for low-mass X-ray binaries (LMXBs) \citep{irwin03}. Assuming this spectrum and 10 photons as a detection threshold, the  source detection sensitivity of the combined observation is $\approx 7\times10^{38} \ \rm{erg \ s^{-1}}$. Since the cluster fills the entire field-of-view (FOV) of the ACIS-S3 detector, we used ``blank-sky'' (http://cxc.harvard.edu/contrib/maxim/acisbg/) observations to subtract the background components. To account for variations in the normalizations of the background, we normalized the background level using the hard band (10-12 keV) count rate ratios. 

Temperature maps were derived using the method of \citet{randall08}. For each temperature map pixel, we extracted a spectrum from a circular region containing a minimum number of $1000$ net counts in the $0.6 - 4.0$ keV band. The resulting spectrum was fitted with an absorbed \textsc{APEC} model using \textsc{XSPEC}, with the abundance allowed to vary.

\begin{figure}[t]
    \leavevmode\epsfxsize=8.5cm\epsfbox{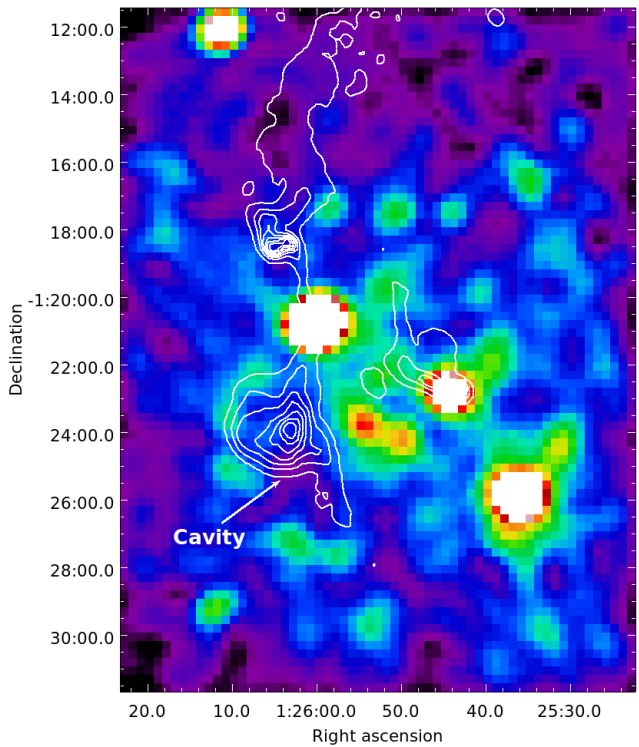}
    \caption{$0.5-1.5$ keV band background subtracted and vignetting corrected \textit{ROSAT} image of the central $324 \times 432$ kpc region of Abell 194 cluster. Overlaid are the intensity levels of the $1.4$ GHz VLA image. It is striking that the southern radio lobe is coincident with a clear cavity in the X-ray surface brightness distribution.}
\label{fig:rosat_image}
\end{figure}

\subsection{\textit{ROSAT}}
\label{sec:rosat}
\textit{ROSAT} observed Abell 194 in a pointed PSPC observation on 1992 July 13 for 23.6 ks  (Fig. \ref{fig:abell194} right panel). Data analysis was performed with standard Xselect tools and a vignetting correction was applied using the provided exposure map. Due to the large FOV, the background was subtracted from the outer regions of the detector which is not affected by emission from Abell 194. To study the diffuse emission from the cluster, the brightest point sources were excluded based on the  \textit{Chandra} source list, with  source cell sizes appropriate for the point spread function of the \textit{ROSAT} PSPC.

\subsection{VLA data}
\label{sec:radio}
To study the properties of Abell 194 in radio wavelengths, we rely on the $1.4$ GHz radio observation  taken by the VLA on 1997 September 16. The observation was performed  in the VLA C configuration, the mean rms noise in the image is $0.151 \ \rm{mJy/beam}$. The radio power was computed as $P_\nu = 4 \pi D_{\nu}^2 S_\nu$, where $D_\nu = D_{L} (1+z)^{-(1+\alpha)/2}$. The bolometric radio luminosity was obtained by integrating the radio flux as 
$$L_{\rm{radio}}=4 \pi D_{L}^2 S_{\nu_0} \int_{\nu_1}^{\nu_2} (\nu / \nu_0)^{- \alpha} d\nu  \rm{,}$$ 
where $\nu_1 = 10 $ MHz and $\nu_2 = 10000 $ MHz and $S_{\nu_0}$ is the VLA flux at $1.4$ GHz.  Additionally, a power law spectrum ($S_{\nu} \propto \nu^{- \alpha}$) is assumed, where $\alpha$ is the  spectral index. The exact value of $\alpha$ is somewhat uncertain: \citet{andernach80}, \citet{kuehr81}, and \citet{spinrad85} give spectral indices of $0.7$, $0.87$, and $0.66$, respectively.  In the further discussion we adopted the mean of these values, namely  $\alpha=0.74$.

\begin{figure}
    \leavevmode\epsfxsize=8.5cm\epsfbox{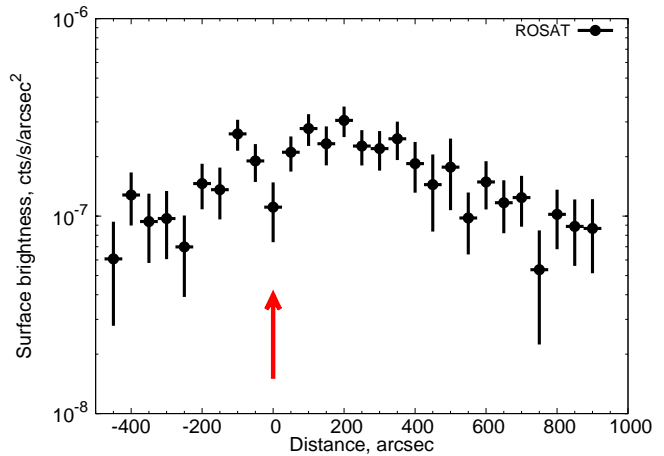}
    \caption{Surface brightness distribution of the $0.5-1.5$ keV band \textit{ROSAT} PSPC image. Background components are subtracted and vignetting correction is applied. On the \textit{x}-axis distance is measured from the  peak of the radio emission and shows the decrement in the X-ray surface brightness (arrow). The values along the  \textit{x}-axis decrease towards east.}
\label{fig:rosat_profile}
\end{figure}

\subsection{Two-Micron All Sky Survey}
To trace the stellar light of the massive early-type galaxies in Abell 194, we used the Two-Micron All Sky Survey (2MASS) Large Galaxy Atlas (LGA) \citep{jarrett03}, which is commonly applied for this purpose \citep[e.g.][]{bogdan10}. Since the 2MASS K-band images of Abell 194  are not background subtracted, we used nearby regions off the galaxies to estimate the background level. To convert counts to luminosity, we assumed that the absolute K-band magnitude of the Sun is $ M_{K,\odot} = 3.28 $.

\section{The X-ray cavity}
\subsection{Detection of the cavity}
In the right panel of Fig. \ref{fig:abell194} we show the $1.4$ GHz VLA image of the central $216\times288$ kpc  region of Abell 194.  The image is dominated by giant radio lobes originating from 3C 40B (WAT), which extend up to $\sim100$ kpc. Note that the southern lobe is within the FOV of the ACIS-S3 detector. On the western side of the image another, much less prominent, jet is observable from 3C 40A (NAT), which is believed to trigger star-formation in Minkowski's Object (Sect. \ref{sec:mo}). In the left panel of Fig. \ref{fig:abell194}, we present a $0.98\arcsec$ Gaussian smoothed $0.3-2$ keV band \textit{Chandra} ACIS-S3 image of the same region. The image was background subtracted and corrected for vignetting as described in Sect. \ref{sec:chandra}. The most striking objects on the image are the three massive early-type galaxies, two of them hosting bright AGN. Additionally, truly diffuse emission is also present, which is concentrated in the region enclosed by NGC547 and NGC541.

The large radio lobes indicate a relatively recent outburst from 3C 40B. In general, jets extending outside from the central AGN inflate radio lobes, thereby pushing the X-ray emitting gas  away and creating bubbles of relativistic plasma. The directly observable consequence of this process is the detection of cavities in the X-ray surface brightness distribution. Therefore in Abell 194 we expect to observe a decrement in the X-ray surface brightness at the position of the radio lobe. To unveil  such a decrement we studied \textit{Chandra}, \textit{ROSAT}, and \textit{XMM-Newton} data. 

First, we studied the $0.3-2$ keV band \textit{Chandra} image to identify the cavity. We selected two circular regions with $1\arcmin$ radii on the southern part of the cluster, marked on the left panel of Fig. \ref{fig:abell194}. One region is centered at the expected position of the cavity (RA: 01h26m05.2s; Dec: -01d23m46.1s), whereas the other is located in the south-western part of the cluster (RA: 01h25m47.0s; Dec: -01d25m15.8s). If an X-ray decrement exists, a significantly lower X-ray flux is expected from the former region. The observed counts from both regions were converted into flux using an optically-thin thermal plasma emission model (\textsc{Mekal} model in \textsc{XSPEC}) with $kT=3.0$ keV, which is the best-fit temperature of the ICM. The obtained flux in the cavity-region is $F_1=(2.95\pm0.15)\times10^{-14} \ \rm{erg \ s^{-1} \ cm^{-2}}$, and is significantly lower than  the neighboring region, $F_2=(3.75\pm0.16)\times10^{-14} \ \rm{erg \ s^{-1} \ cm^{-2}}$. This suggests the existence of an X-ray cavity coincident with the southern 3C 40B radio lobe. 

Second, to confirm the existence of the cavity, we  investigated the available \textit{ROSAT} PSPC data, whose major advantages are its  low background level and its large FOV. In Fig. \ref{fig:rosat_image}   a $0.5-1.5$ keV band vignetting corrected and background subtracted \textit{ROSAT} image is shown and overlaid are the $1.4$ GHz VLA contours. The image shows that the hot ICM fills the cluster, and reveals an X-ray decrement, coincident with the southern radio lobe. To further illustrate the presence of the cavity we show the $0.5-1.5$ keV band X-ray surface brightness distribution in Fig. \ref{fig:rosat_profile}. The profile was obtained using rectangular boxes oriented in the east-west direction, whose extent were $50\arcsec \times 125\arcsec$.  The region at $0\arcsec$ distance is  centered in the peak of the radio emission, i.e. in the center of the X-ray cavity. Note that the profile is corrected for vignetting and background is subtracted, furthermore it represents  the diffuse emission since point sources were excluded as described in Sect. \ref{sec:rosat}. In agreement with our previous observations, Fig. \ref{fig:rosat_profile} reveals a sharp drop at the expected position of the X-ray cavity, and shows a relatively smooth gas distribution elsewhere. 

Finally, as a consistency check we also investigated whether the available \textit{XMM-Newton} EPIC data shows an evidence of the X-ray cavity. The data was prepared as described in \citet{bogdan08}. The same circular regions with $1\arcmin$ radii were used to describe the cavity and the neighboring regions as for the \textit{Chandra} data (Fig. \ref{fig:abell194} left panel). The observed counts were converted to flux using a \textsc{Mekal} model with $kT=3.0$ keV. The obtained flux in the cavity-region is $F_1=(2.8\pm0.4)\times10^{-14} \ \rm{erg \ s^{-1} \ cm^{-2}}$, and is significantly lower than  the neighboring region, $F_2=(4.0\pm0.5)\times10^{-14} \ \rm{erg \ s^{-1} \ cm^{-2}}$. The observed flux is in good agreement with that obtained by \textit{Chandra}.  Although the difference between the X-ray flux in the  cavity and neighboring region is  statistically significant, the significance is somewhat smaller due to the larger error bars, caused by the higher and less stable background of \textit{XMM-Newton}'s EPIC  detectors.

Thus,  \textit{Chandra}, \textit{ROSAT}, and \textit{XMM-Newton} results demonstrate the existence of a highly statistically significant X-ray cavity in Abell 194, produced by the giant southern radio lobe arising from 3C 40B.

\subsection{Cavity power}
\label{sec:cavity}
Measuring the cavity power ($P_{\rm{cav}}$) offers a straightforward estimate of the energy injected to the X-ray emitting gas by the AGN outburst. $P_{\rm{cav}}$ can be estimated by computing the minimal energy required to inflate the cavity and the age of the cavity. The former is obtained by measuring   the cavity enthalpy as $H=4pV$ (for relativistic plasma), whereas the latter is calculated as  the sound crossing time. However, the calculation of $P_{\rm{cav}}$ involves a number of uncertainties (see below), hence the obtained cavity power should be considered as a sensible estimate. 

Based on the VLA data, we describe the detected cavity in Abell 194  with a circular region with a radius of $34$ kpc ($\approx1.57\arcmin$) centered on the coordinates: RA: 01h26m05.2s, Dec$\rm{:-}$01d23m46.1s.  The distribution of the cavity is assumed to be spherically symmetric. Since the ICM is not strongly peaked in Abell 194, we used the average ICM properties  within the \textit{Chandra} FOV to obtain the physical parameters of the gas. Therefore we selected a  circular region with $4\arcmin$ radius centered in the center of the ACIS-S3 detector. Note that resolved point sources were excluded along with three circular regions around NGC541, NGC545, and NGC547 with radii of $15\arcsec$, $13\arcsec$, and $13\arcsec$, respectively. The spectra of the $4\arcmin$  large region was described with an optically-thin thermal plasma emission model (\textsc{Mekal} model in \textsc{XSPEC}) with a best-fit temperature of $kT=3.0\pm0.3$ keV ($90\%$ confidence interval). From the normalization of the model the emission measure is $\int n_e n_H dV = 3.0 \times10^{65} \ \rm{cm^{-3}}$, implying an average density of $n=1.9 \times 10^{-3} \ \rm{cm^{-3}}$. Thus, the average pressure is $p=1.9 n_e kT = 1.7 \times 10^{-11} \ \rm{erg \ cm^{-3}}$. Using these parameters and computing the volume of the cavity region, the total AGN work is deduced as $4pV=3.3 \times 10^{59}$ erg. 

To compute the cavity power from the $4pV$ work, the age of the cavity has to be estimated. To apply a realistic expansion velocity, we consider two observational facts. On the one hand the lack of a shock indicates that  the expansion of the lobe is not very supersonic, on the other the lobe is not strongly deformed by the buoyancy force and hence the expansion is also not very subsonic. Therefore, a somewhat uncertain but reasonable assumption is that the radio lobes propagate with the sound speed, which, in a $kT=3.0$ keV plasma, is  $c_s=\sqrt{(\gamma kT)/(\mu m_H})=875 \ \rm{km \ s^{-1}}$ by adopting $\gamma=5/3$ and $\mu=0.62$. Assuming that the lobes are in the plane of the sky, the center of the cavity is approximately $d=71$ kpc ($\approx3.3\arcmin$) from 3C 40B, hence the cavity age is $t= d/c_s= 7.9 \times 10^7$ years.  Based on these,  the approximate cavity power of  the southern  cavity is  $P_{cav}=4pV/t=1.3 \times 10^{44} \ \rm{erg \ s^{-1}}$.

\begin{figure*}[t]
  \begin{center}
    \leavevmode
      \epsfxsize=8.5cm\epsfbox{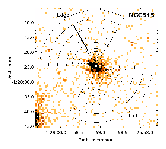}
\hspace{0.5cm} 
      \epsfxsize=8.5cm\epsfbox{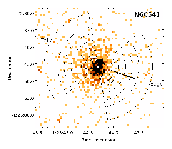}
     \epsfxsize=8.5cm\epsfbox{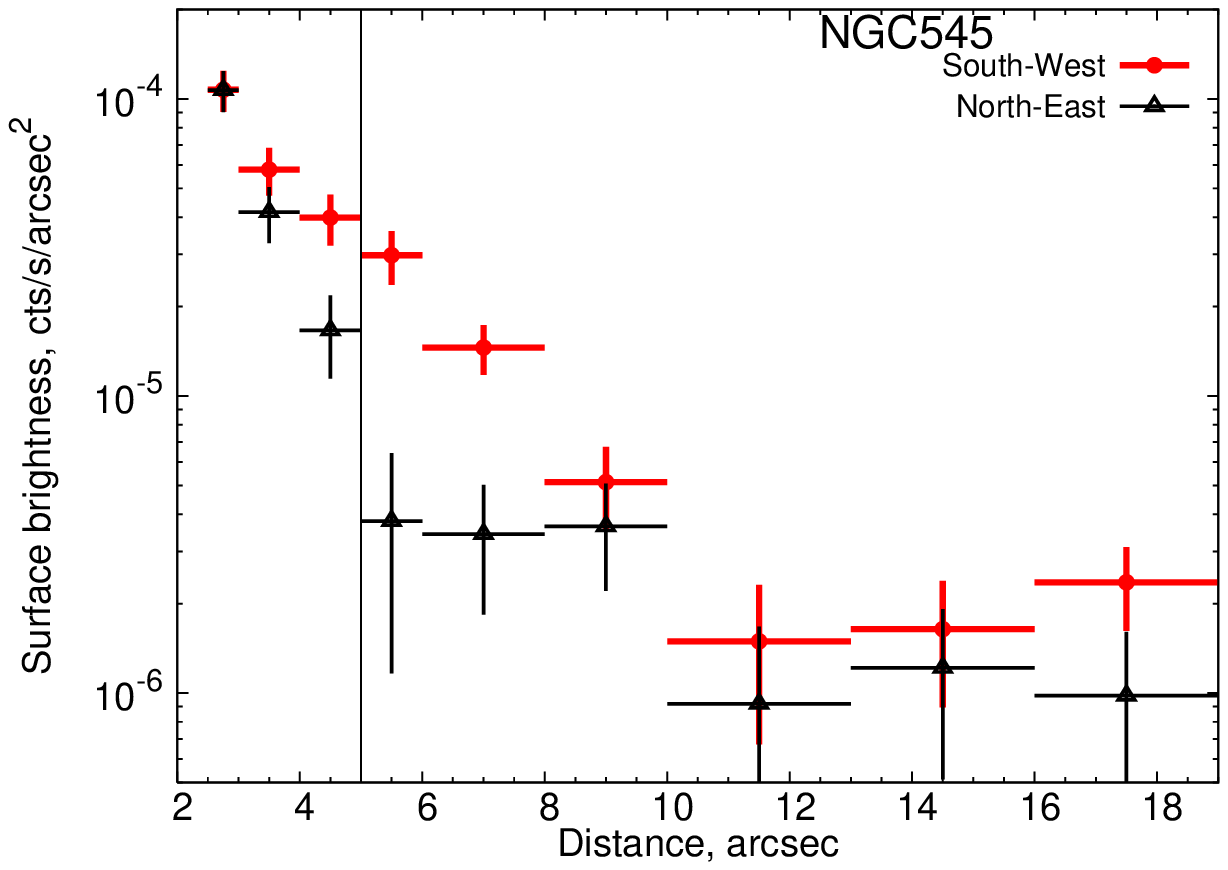}
\hspace{0.5cm} 
      \epsfxsize=8.5cm\epsfbox{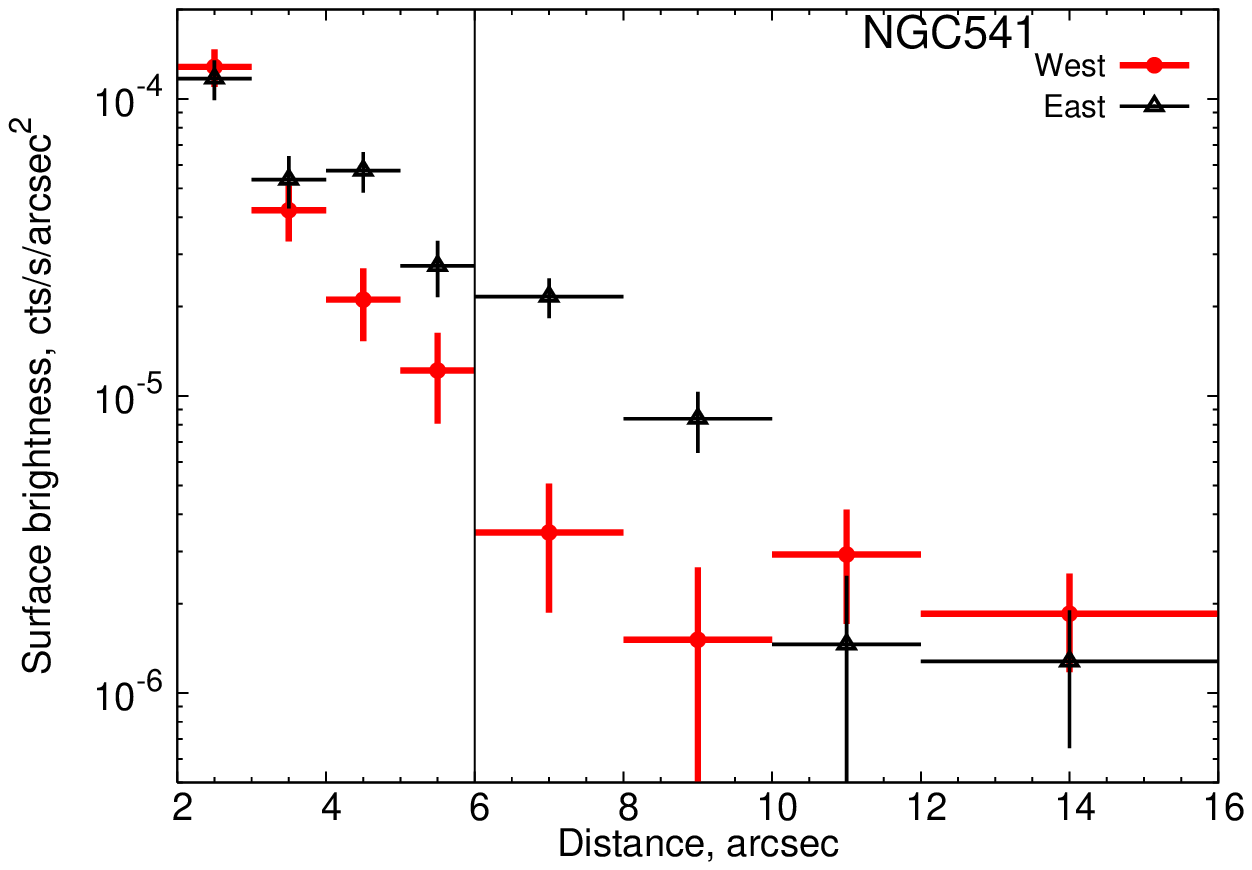}
      \caption{\textit{Top:} $0.3-2$ keV \textit{Chandra} images of NGC545 (left panel) and NGC541 (right). Overplotted are the regions used to obtain surface brightness profiles shown in the bottom panels. The surface brightness edges are located on the north-eastern and western sides of NGC545 and NGC541, respectively.  The arrows show the approximate direction of motion of the galaxies (Sect. \ref{sec:dynamics}). Note that the bend in NAT in NGC541 is consistent with the direction of motion implied by the edge. The center of galaxies is marked with the cross. \textit{Bottom}: X-ray surface brightness distributions in the $0.3-2$ keV energy range around NGC545 and NGC541 using the regions shown in the upper panels. Filled circles (red) represent the south-western and western sides, whereas  triangles (black) illustrate the north-eastern and eastern sides of NGC545 and NGC541, respectively. The thin vertical lines mark the approximate position of the surface brightness edges.}
\vspace{0.5cm}
     \label{fig:edge}
  \end{center}
\end{figure*}

\subsection{Comparison with other radio induced cavities}
During the operation of \textit{Chandra} the existence of radio induced X-ray cavities has been shown in many galaxies and galaxy clusters. The correlation between the cavity power ($P_{\rm{cav}}$) and the corresponding radio luminosity ($L_{\rm{radio}}$) was studied  in a broad sample of systems by \citet{birzan04,birzan08}.  The best-fit relation for the total source luminosity, given by \citet{birzan08} is:
$$\log P_{\rm{cav}} = (0.48 \pm 0.07) \log L_{\rm{radio}} + (2.32\pm0.09) \rm{,}$$ 
where $L_{\rm{radio}}$ represents the bolometric radio luminosity. In this relation both $P_{\rm{cav}}$ and $L_{\rm{radio}}$ are in units of $10^{42} \ \rm{erg \ s^{-1}}$. Since the detected cavity in Abell 194 is similar to those investigated earlier \citep{birzan04,birzan08}, we examine if its properties fit this relation. 

The bolometric radio luminosity was computed from the 1.4 GHz VLA data as described in Sect. \ref{sec:radio}, which  yielded $L_{\rm{radio}}= 1.7 \times 10^{41} \ \rm{erg \ s^{-1}}$, whereas the cavity power is $P_{cav}=1.3 \times 10^{44} \ \rm{erg \ s^{-1}}$ (Sect. \ref{sec:cavity}). Substituting the observed radio luminosity in the \citet{birzan08} relation, we find that the corresponding predicted cavity power is about $8.9 \times 10^{43} \ \rm{erg \ s^{-1}}$.  This value is similar albeit somewhat lower than the observed cavity power in Abell 194. Considering the uncertainties in the estimation of $P_{cav}$, and the dispersion in the relation given by \citet{birzan08},  we conclude that the detected cavity in Abell 194 is in  good agreement with those  obtained in other galaxies and clusters of galaxies.

\section{Cold fronts in NGC545 and NGC541 and large scale dynamics in Abell 194}
\label{sec:merger}
\subsection{Cold fronts in NGC545 and NGC541}
\label{sec:coldfront}
Two galaxies in Abell 194, namely NGC545 and NGC541, exhibit sharp surface brightness edges and extended tails in the soft band  \textit{Chandra} images (Fig. \ref{fig:edge} upper panels). The edges are located on the north-eastern and western side of NGC545 and NGC541, respectively. To demonstrate the existence  of the edges and tails, we extracted surface brightness profiles of the galaxies using the regions displayed in the upper panels of Fig. \ref{fig:edge}. The  background subtracted $0.3-2$ keV band profiles are depicted in the lower panels of  Fig. \ref{fig:edge}. As suggested by the images, the X-ray surface brightness drops at $\sim5\arcsec$ on the north-eastern and western sides of NGC545 and NGC541, whereas on the opposite sides it extends beyond $\sim10\arcsec$ radius. 

To further study the X-ray emission inside and outside the edges, we extracted X-ray energy spectra from both regions. Due to the relatively low number of counts, only one region  could be investigated  inside and outside the edges.  The former was represented by circular regions centered on each galaxy with  $5\arcsec$ ($1.80$ kpc) and $6\arcsec$ ($2.16$ kpc) radii for NGC545 and NGC541, respectively. The soft emission was described with a \textsc{Mekal} model and the abundances were fixed at $0.5$ solar \citep{anders89}. The   best-fit temperatures were $0.63\pm0.04$ keV and $0.55\pm0.05$ keV for NGC545 and NGC541, respectively. Regions outside the edges were represented by similar wedges to those  depicted in Fig. \ref{fig:edge}, but their radial extents were $5\arcsec-75\arcsec$ ($1.8-27$ kpc) and $6\arcsec-75\arcsec$ ($2.16-75$ kpc) for NGC545 and NGC541, respectively. The regions were oriented towards the north-eastern side of NGC545 and the western side of NGC541. The  spectra demonstrated the presence of  ICM, with emission described  with a \textsc{Mekal} model with abundances fixed at $0.3$ solar. The best-fit temperatures were $3.2^{+0.9}_{-0.7}$ keV in NGC545 and $2.8^{+0.8}_{-0.6}$ keV in NGC541. 

The observed sharp surface brightness  edges in NGC545 and NGC541, and  the temperature difference between the regions inside and outside the edges suggest that these features are contact discontinuities or  cold fronts  \citep{vikhlinin01,machacek05,markevitch07}.

\subsection{Large scale dynamics of Abell 194}
\label{sec:dynamics}
The detection of cold fronts in NGC545 and NGC541 permits us to compute their velocity, and hence map the large scale dynamics of Abell 194. Ideally, their velocities  can be directly calculated from X-ray observations using the ratio of  the thermal pressures at the stagnation point ($p_0$) and in the free stream ($p_1$) \citep{vikhlinin01}. However, due to the relatively low number of counts a number of assumptions are involved, and hence only  crude estimates can be given. On the one hand 
$p_0$ cannot be constrained across the   surface brightness edges, instead we used the average pressure in the galaxy;
on the other outside the cold fronts $p_1$ can only be measured in a relatively large region. Thus, we assume that no significant temperature and pressure gradients are present inside and outside the galaxies. Additionally, the poorly constrained abundances further limit the accuracy of the estimates. 

The pressure ratios were estimated using the regions described in Sect. \ref{sec:coldfront}.  We derived  $p_0/p_1=2.1^{+0.7}_{-0.9} $ for NGC545 and  $p_0/p_1=1.8^{+0.5}_{-0.7}$  for NGC541, which translate  to Mach numbers of $M=1.0^{+0.3}_{-0.5} $ for NGC545 and $M=0.9^{+0.2}_{-0.5}$  for NGC541. The sound speed in  $3$ keV gas is $c_s=875 \ \rm{km \ s^{-1}}$, hence the  Mach numbers correspond to velocities of $875^{+263}_{-438}   \ \rm{km \ s^{-1}}$ and $788^{+175}_{-438}   \ \rm{km \ s^{-1}}$ for NGC545 and NGC541, respectively.

\begin{figure}[t]
\begin{center}
    \leavevmode\epsfxsize=8.5cm\epsfbox{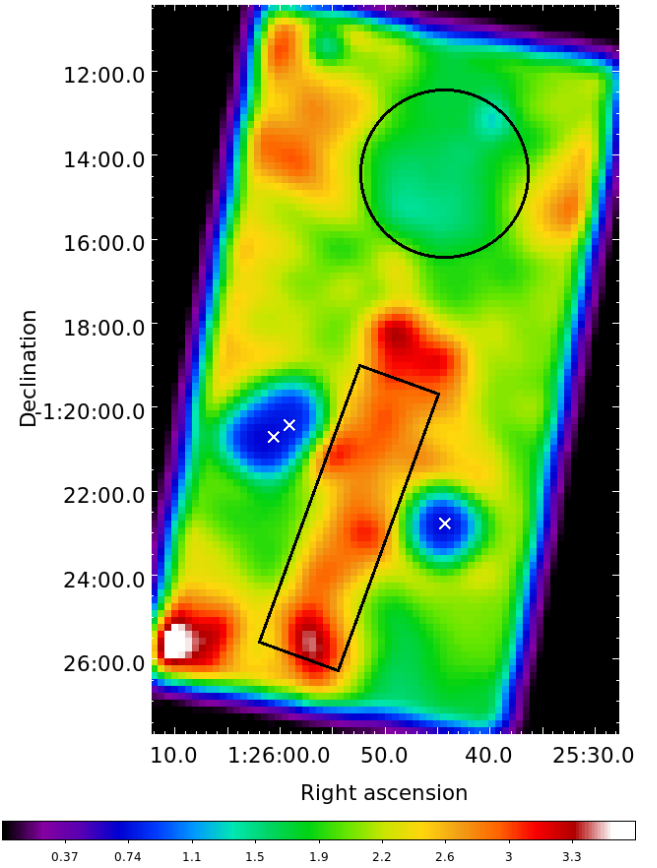}
    \caption{Temperature map of the central regions of Abell 194 derived from ACIS-S3 and ACIS-S2 data. The color-bar gives the temperature in keV. The crosses mark the positions of the bright cluster galaxies: on the eastern side of the cluster are located NGC547 (southern cross) and NGC545 (northern cross), whereas NGC541 is on the western side. The overplotted regions were used for spectral extraction. Note the increased ICM temperature between the bright cluster galaxies. }
\label{fig:tmap}
\end{center}
\end{figure}

Based on the velocities of NGC545 and NGC541 and their  radial velocities,  the orientation of the galaxy motions can be deduced. The radial velocities of NGC545 and NGC541 are $ 5338 \ \rm{km \ s^{-1}}$ and $ 5422 \ \rm{km \ s^{-1}}$ (NED), implying a velocity of $ -96 \ \rm{km \ s^{-1}}$ and $ 26 \ \rm{km \ s^{-1}}$ relative to the mean radial velocity of Abell 194 ($ 5396 \ \rm{km \ s^{-1}}$). Thus, we conclude that both NGC545 and NGC541 are moving essentially in the  plane of the sky, since the direction of their motion differs $\lesssim10\degr$ from it. 

\begin{figure*}[t]
  \begin{center}
    \leavevmode
      \epsfxsize=8.5cm\epsfbox{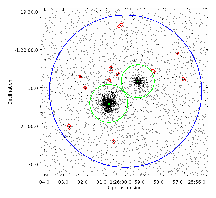}
\hspace{0.5cm} 
      \epsfxsize=8.5cm\epsfbox{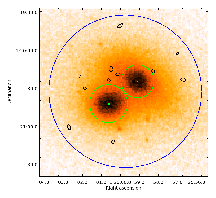}
     \epsfxsize=8.5cm\epsfbox{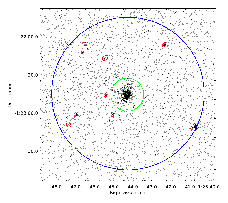}
\hspace{0.5cm} 
      \epsfxsize=8.5cm\epsfbox{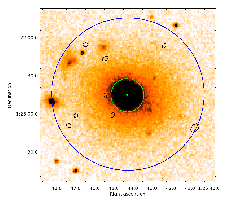}
      \caption{\textit{Left:} $0.5-8$ keV band \textit{Chandra} image of the NGC547, NGC545 regions (top panel) and NGC541 region (bottom panel). The large circle (blue) represents the $1\arcmin$ regions around the galaxies, whereas the small (green) circles ($15\arcsec$ for NGC547 and $13\arcsec$ for NGC545 and NGC541) show the regions excluded from the analyses. The detected point sources having at least $10$ counts are also encircled. The center of galaxies is marked with the cross. \textit{Right:} DSS $R$-band images of the same regions as on the left panel. The meaning of the circles is the same as before. Note, that only one X-ray source on the north-eastern part of NGC541 (bottom) has an optical counterpart.}
\vspace{0.5cm}
     \label{fig:point_sources}
  \end{center}
\end{figure*}

The observed cold fronts and the estimated transonic velocities of NGC545 and NGC541  suggest that they are falling through the cluster.  The gas distribution in NGC547 appears to be symmetric and  does not reveal any surface brightness edges or tails, indicating its relatively low velocity. Indeed, from the bending of the radio  jet emanating from 3C 40B \citet{sakelliou08} estimated that NGC547 moves with subsonic velocity, $v_{\rm{NGC547}}\le300 \ \rm{km \ s^{-1}}$. Additionally, the radial velocities of NGC547, NGC545, and NGC541 are very similar to each other: $ 5468 \ \rm{km \ s^{-1}}$, $ 5338 \ \rm{km \ s^{-1}}$, and  $ 5422 \ \rm{km \ s^{-1}}$, respectively. The minor differences in the radial velocities and the small projected distances between these galaxies suggest that they compose a gravitationally bound system. Thus,  the center of Abell 194 is presumably NGC547, and NGC545 and NGC541 are falling through the  cluster, implying that Abell 194 is undergoing a significant cluster merger event. 

The temperature map of the central regions of Abell 194 is depicted in Fig. \ref{fig:tmap}. Interestingly, the temperature map reveals a hot elongated region between NGC547/545 and NGC541, and cooler temperature outside this region. To verify the  temperature difference, the spectra of two regions were extracted (Fig. \ref{fig:tmap}).  One of them represents the hot region between the bright cluster galaxies, whereas the other corresponds to the cooler region north of these galaxies. The obtained best-fit temperatures were $kT = 3.2^{+0.8}_{-0.6}$ keV and  $kT = 1.4^{+0.5}_{-0.3}$ keV ($90\%$ confidence intervals), which indicate a statistically significant difference. The higher temperature between NGC547/545 and NGC541 is most likely caused by a reverse merger shock \citep{markevitch01}. Note, that the forward shock cannot be unambiguously observed since on the one hand it is likely partly outside the FOV, on the other it may be too faint due to the low ICM density. We thus conclude that the detection of hot ICM between the bright cluster galaxies is consistent with the picture of the galaxy dynamics in Abell 194. 

The non-detection of large scale cold fronts in Abell 194 indicate that the merger is not between two roughly equal mass systems in the first pass stage. Instead,  the galaxies may have already undergone more than one pass, during which most of the cold gas was stripped from them. In agreement with this, the optical/X-ray bridge between NGC547/NGC545 and NGC541 \citep{croft06} also indicates past and/or recent interactions between them, implying that the potential  probably extends to the western side of the cluster as well.

\section{Overdensity of bright sources around massive early-type galaxies in Abell 194}
\subsection{Predicted number of LMXBs and CXB sources}
A surprising result of our analysis is that around the massive early-type galaxies of Abell 194 (NGC541, NGC545, NGC547) an unusually high number of bright resolved sources are observed (Fig. \ref{fig:point_sources}). To study the number of sources around these galaxies, two  circular regions were selected with $1\arcmin$ radius: the first is centered  between NGC547 and NGC545 (henceforth region $A$) whereas the second is centered on NGC541 (henceforth region $B$). The central parts ($15\arcsec$ for NGC547 and $13\arcsec$ for NC541 and NGC545) of galaxies were excluded from the analysis since their bright X-ray coronae notably reduce the point source detection sensitivity. Considering only sources with at least 10 counts in the $0.5-8$ keV energy range, in regions $A$ and $B$ $11$ and $9$ sources were detected, respectively. 

The most likely origins of the detected sources are:  first, they could be low-mass X-ray binaries (LMXBs) associated with the galaxies; second, they may be resolved cosmic X-ray background (CXB) sources. 

The number of resolved LMXBs above a certain  sensitivity limit can be determined using their average luminosity function \citep{gilfanov04}. In the selected regions the source detection sensitivity is $\approx7.2 \times10^{38} \ \rm{erg \ s^{-1}}$. The corresponding K-band luminosities are $2.4 \times 10^{11} \ \rm{L_{K,\odot}}$ and $1.5 \times 10^{11} \ \rm{L_{K,\odot}}$ in regions $A$ and $B$, respectively. According to the LMXB luminosity function we predict $0.8$ and $0.5$ resolved LMXBs above the detection threshold in regions $A$ and $B$, respectively. Thus, only $\approx6\%$ of the resolved sources can be explained by LMXBs. 

The number of predicted CXB sources were determined using two independent approaches. As a first method, we used the $\log N - \log S$ function of \citet{moretti03}. We converted the observed $0.5-8$ keV band sensitivity limit  ($1.14 \times 10^{-15} \ \rm{erg \ s^{-1} \ cm^{-2}} $) to the $2-10$ keV band assuming a power law spectral model with $\Gamma=1.4$ and Galactic column density, which results in the sensitivity limit of $1.03 \times 10^{-15} \ \rm{erg \ s^{-1} \ cm^{-2}} $. Based on this value, the surface area of regions $A$ and $B$, and  the $\log N - \log S$ function, we predicted $1.9$ and $2.0$ sources in $A$ and $B$ regions, respectively. As a second method, we used the field of Abell 194 to determine the average surface density of resolved CXB sources. Using a circular region with $3\arcmin$ radii, centered in the CCD, $17$ point sources were observed.   Note that in this computation the sources located within regions $A$ and $B$ are excluded, along with the point source associated with the elliptical galaxy CGCG 385-127. Based on the surface area of  regions $A$ and $B$, we expect $2.0$ and $2.1$ sources within these regions. We stress that the predicted numbers of CXB sources obtained with the two methods are in excellent agreement with each other. Therefore no more than $\approx20\%$ of the detected sources could arise from resolved CXB sources. 

Based on these, in the first instance, we find that LMXBs and CXB sources altogether can be responsible for $\approx5.4$ sources, whereas the observed number is $20$, implying a $>5\sigma$ excess of point sources.  We mention that \citet{hudaverdi06} reported an excess of sources in Abell 194 (and Abell 1060) in the luminosity range of $10^{39.6} \le L_X \le 10^{41.4} \ \rm{erg \ s^{-1}}$ over the full \textit{XMM-Newton} FOV. They suggested that these sources, many identified with cluster member galaxies, are AGN within cluster-member galaxies.  The source excess we report has a very different origin since these sources appear to be associated with the massive early-type galaxies in Abell 194, and are fainter, typically $L_X \sim 10^{39} \ \rm{erg \ s^{-1}}$, than those studied by \citet{hudaverdi06}.

\subsection{Possible origin of the source overdensity}
The normalization of the LMXB luminosity function exhibits a scatter of about a factor of two \citep[e.g.][]{gilfanov04}. Although such variations cannot explain the total number of excess sources, it may be responsible in part. The stable shape of the luminosity function implies that if its bright end normalization is higher, then its faint end must also have an elevated level. Thus, we can  scale the average luminosity function using the $2-8$ keV band flux from the unresolved emission in regions $A$ and $B$. The  derived total hard band unresolved X-ray luminosity of $L_X=3.9 \times 10^{40} \ \rm{erg \ s^{-1}}$ corresponds to $L_X/L_K=1.0\times10^{29} \ \rm{erg \ s^{-1} \ L_{K,\odot}^{-1} }$. Below the $7.2 \times10^{38} \ \rm{erg \ s^{-1}}$ sensitivity limit, the average  LMXB luminosity function \citep{gilfanov04} predicts $L_X/L_K=5.9\times10^{28} \ \rm{erg \ s^{-1} \ L_{K,\odot}^{-1} }$ in the $2-8$ keV energy band, assuming an average power law LMXB spectrum with slope of $\Gamma=1.56$ and Galactic column density \citep{irwin03}. Thus, the luminosity function of LMXBs has to be scaled up by $\approx70\%$. Therefore the total predicted number of LMXBs is $2.2$ in regions $A$ and $B$. Taking this as the expectation value, the total number of predicted LMXBs and CXBs is $6.3$, whereas $20$ sources are observed. 

An additional major uncertainty in the average LMXB luminosity function is the relatively low number of  bright ($L_X \gtrsim8 \times10^{38} \ \rm{erg \ s^{-1}} $) sources in the sample of \citet{gilfanov04}. This implies large systematic uncertainties at the bright end of the luminosity function. Indeed, according to \citet{gilfanov04} the number of birght LMXBs above  $L_X\gtrsim8 \times10^{38} \ \rm{erg \ s^{-1}} $ may be factor of $\sim4$ times higher within  $90\%$ confidence interval due to systematic errors. Taking this uncertainty into account, it is feasible that significantly more, altogether $8-9$, resolved sources are bright LMXBs in regions $A$ and $B$. Therefore LMXBs and CXB sources may account for up to $11$ sources of the $20$ detected.  

Since no significant star-formation is associated with early-type galaxies, it is implausible that the additional population of point sources are high-mass X-ray binaries (HMXBs). To illustrate this point we estimate the star-formation rate (SFR) required to produce 9 HMXBs brigher than $7.2\times10^{38} \ \rm{erg \ s^{-1}}$. Relying on the average HMXB luminosity function from \citet{grimm03} and using the normalization by \citet{shtykovskiy05} the required SFR is $\sim18 \ \rm{M_{\odot}/yr}$. As this value is  consistent with those observed for starburst galaxies, we conclude that the importance of HMXBs  is negligible. 

Since the galaxies we consider are in a cluster, it is possible that (at least some of) the X-ray sources are associated with  galaxies within the cluster, alternatively foreground stars may play a role. To identify such galaxies we cross-checked  the coordinates of all X-ray sources in NED. Additionally we visually inspected Hubble Space Telescope and DSS images (Fig. \ref{fig:point_sources} right panel) to search for optical counterparts of X-ray sources. We found only one X-ray source around NGC 541 (RA: 01h25m46.66s; Dec: -01d22m12.97s) which  has an optical counterpart. Excluding this source the observed number of X-ray sources is $19$, whereas LMXBs and CXB sources may account for up to $11$ sources.

As is clear from Fig. \ref{fig:point_sources}, NGC547, NGC545, and NGC541 have very luminous X-ray coronae extending to large radii, within which  a notable fraction of the detected sources are located. As our experience shows, it is feasible that some (unknown) fraction of the resolved sources are not pointlike compact sources, but are features of the shock-heated gas distribution. Obviously, such features in the gas distribution would further decrease the discrepancy between the observed and predicted number of resolved pointlike sources.  However, taking the depth of the \textit{Chandra} observations and the distance of Abell 194, with the currently available exposures  this question cannot be  resolved. 

We conclude that in regions $A$ and $B$ a $\gtrsim2\sigma$ excess of point sources is detected, which may (partly) be due to unresolved features of the shocked X-ray gas and/or an excess population of bright sources. 

\section{Minkowski's Object}
\label{sec:mo}
\subsection{Origin of the X-ray emission}
Minkowski's Object (MO) \citep{minkowski58} lies along the jet emanating from NGC541 (3C 40A), and is believed to be a classic example of jet-triggered star-formation in the local Universe. MO was  studied by \citet{breugel85} and \citet{brodie85}, who suggested that the observed emission originated in a starburst. More recently  a multi-wavelength study by \citet{croft06} found that the stellar population of MO is dominated by a $7.5\times10^6$ years old, $1.9\times10^7 \ \rm{M_{\odot}}$ instantaneous burst, with a current SFR of $ 0.52 \ \rm{M_{\odot}/yr} $.

In actively star-forming galaxies the dominant fraction of the observed X-ray emission is a combination of the population of HMXBs and the diffuse hot ISM \citep[e.g.][]{grimm03,bogdan11}. However, due to the relatively high source detection sensitivity ($\approx7\times10^{38} \ \rm{erg \ s^{-1}}$) in MO, most of the HMXBs are not resolved. Indeed, in a circular region with $15\arcsec$  radius around MO  we detected only one X-ray source with 8 counts ($L_X \approx 5.6 \times 10^{38} \ \rm{erg \ s^{-1}}$). Thus, in MO the various X-ray emitting components cannot be separated from each other since they all contribute to the unresolved emission. Therefore we can only place upper limits on the observed X-ray emission by assuming that it originates either from unresolved HMXBs or from hot ISM. 

First, we assume that the total X-ray luminosity is due to the population of unresolved HMXBs. The observed source counts from the selected $15\arcsec$ circular region in the $0.5-8$ keV band  correspond to a luminosity of $(2.3\pm0.4)\times10^{39} \ \rm{erg \ s^{-1}}$ assuming a typical HMXB power law spectrum with slope of $\Gamma=2$ and Galactic column density. Converting this luminosity into the $2-10$ keV band, and taking into account the SFR in MO \citep{croft06}, we obtained  $L_X/$SFR$=(2.7\pm0.4)\times10^{39} \ \mathrm{(erg/s)/(M_{\odot}/yr)}$. This value is in good agreement with that observed for HMXBs \citep{grimm03, shtykovskiy05}, thereby suggesting that at least part of the observed X-ray emission originates from the population of HMXBs.

Second, we assume that the observed X-ray emission is due to hot ISM. Since the observed number of counts is not enough to perform a spectral fit, we adopt a gas temperature of $kT=1$ keV and $0.4$ Solar abundance. Applying this conversion the observed counts correspond to a luminosity of $(1.0\pm0.2)\times10^{39} \ \rm{erg \ s^{-1}}$. Assuming a spherically symmetric gas distribution within the selected $15\arcsec$ circular region, the estimated average gas density is $4.4 \times 10^{-3} \ \rm{cm^{-3}}$ and the total gas mass in MO is $7 \times 10^{7} \ \rm{M_{\odot}}$. Note that the upper limit on the gas mass is  $\sim4$ times higher than the stellar mass in MO \citep{croft06}. Therefore it is unlikely that the hot ISM  dominates the observed X-ray emission. Additionally, if the X-ray emission in  MO is dominated by hot ISM,  the associated  thermal energy  is on the order of a few times $10^{56} \ \rm{erg}$, the exact value depending on the gas temperature. To inject this large an  amount of energy  into the hot ISM, a few times $10^5$ Type Ia Supernova (SN Ia) explosions are required. On a timescale of $7.5\times10^6$ years, this implies a SN Ia rate of few times $10^{-2} \ \rm{yr^{-1}}$, being $1-2$ orders of magnitude higher than the predicted SN Ia rate in MO based on its SFR \citep{sullivan06,croft06}. Therefore SNe Ia cannot heat $7 \times 10^{7} \ \rm{M_{\odot}}$ gas to  X-ray temperatures, but only a small fraction of it.  Thus, we conclude, that the X-ray emission in MO is either dominated by HMXBs or the jet interaction with the ambient ISM plays a role in heating the gas. 

\subsection{An alternative origin of the star formation in Minkowski's Object}
Although it is widely believed that the star-formation in MO is triggered by the jet emanating from NGC541 (3C 40A), based on our results on the dynamics of Abell 194 we explore an alternate origin of MO. In Sect. \ref{sec:merger} we demonstrated that NGC541 is falling through the cluster and its direction of movement is pointing towards the west. Interestingly,  its motion approximately points back to NGC547/NGC545 and also crosses the position of  MO. This  suggests that the active star-formation in MO could be triggered by its interaction with NGC541. 

The projected distance of MO from NGC541 is $\sim19$ kpc, and the velocity of NGC541 is $788^{+175}_{-438} \ \rm{km \ s^{-1}}$ (Sect. \ref{sec:dynamics}). Assuming that the motion takes place in the plane of the sky, and considering only the velocity of NGC541, the interaction between NGC541 and MO could have happened $(1.9-5.3)\times10^7$ years ago. The age of MO can be estimated by different methods: on the one hand \citet{croft06}  obtained a stellar age of $7.5\times10^6$ years based on spectral energy distribution models; on the other if constant SFR  ($ 0.52 \ \rm{M_{\odot}/yr} $) is assumed and the total  stellar mass of MO ($1.9\times10^7 \ \rm{M_{\odot}}$) is considered, an average age of $\approx3.7\times10^7$ years is computed. Thus, the estimated age of MO is $(0.75-3.7)\times 10^7$ years, which  is in approximate agreement with the time range of the possible interaction between MO and NGC541. 

Thus, it is feasible that the interaction between a preexisting gas-rich dwarf galaxy (now observed as MO) and NGC541 triggered a starburst event, which largely rejuvenated the stellar population of MO.

\section{Conclusions}
We investigated the poor cluster,  Abell 194, based on \textit{Chandra} and \textit{ROSAT} observations. The main results of our work are outlined below. \\

(i) We demonstrated the existence of an X-ray cavity at the southern part of Abell 194 formed by the southern radio lobe arising from 3C 40B. Using the parameters of the hot ICM, the extent and location of the cavity, we estimated the total work of the AGN, $3.3\times 10^{59} \ \rm{erg}$, the age of the cavity, $t=7.9\times10^7$ years, and the total cavity power, $P_{cav}=1.3 \times 10^{44} \ \rm{erg \ s^{-1}}$.

(ii) We detected sharp surface brightness edges, identified as merger cold fronts, and extended tails in the \textit{Chandra} images of NGC545 and NGC541. We estimated that NGC545 and NGC541 are moving with transonic velocities, and their motion is oriented approximately in the plane of the sky. Based on these and earlier observations, we concluded that NGC547, associated with 3C 40B, lies at the center of Abell 194 and NGC545 and NGC541 are falling through the cluster. 

(iii) A $\gtrsim2\sigma$ excess of point sources were detected around the massive early-type galaxies in Abell 194. The additional sources may be (partly) due to unresolved features of the shocked X-ray gas in the coronae of the  galaxies and/or an excess population of bright sources.

(iv) We also investigated the X-ray emission from Minkowski's Object, and concluded that it is most likely dominated by the population of HMXBs rather than by  hot diffuse ISM. Additionally, in the view of the galaxy dynamics in Abell 194, we explored the possibility that the starburst in Minkowski's Object was triggered by the interaction of a preexisting gas-rich dwarf galaxy and NGC541. 

\bigskip
\begin{small}
\noindent
\textit{Acknowledgements.}
The authors thank Martin Hardcastle for critical reading of the manuscript and the anonymus referee for helpful and constructive comments. This research has made use of \textit{Chandra} archival data provided by the \textit{Chandra} X-ray Center. The publication makes use of software provided by the \textit{Chandra} X-ray Center (CXC) in the application package CIAO. This publication makes use of data products from the Two Micron All Sky Survey, which is a joint project of the University of Massachusetts and the Infrared Processing and Analysis Center/California Institute of Technology, funded by the National Aeronautics and Space Administration and the National Science Foundation. The Digitized Sky Survey (DSS) and the NASA/IPAC Extragalactic Database (NED) have been used. The VLA (Very Large Array) is a facility of the National Radio Astronomy Observatory (NRAO). The NRAO is a facility of the National Science Foundation operated under cooperative agreement by Associated Universities, Inc. WF and CJ acknowledge support from the Smithsonian Institution. The financial support for this work was partially provided by the Chandra X-ray Center through NASA contract NAS8-03060, and the Smithsonian Institution.
\end{small}

\end{document}